\documentclass[final]{svjour3}
\smartqed  
\usepackage{graphicx}
\usepackage{natbib}

\journalname{J Econ Interact Coord (2009) 4:221–-235}

\begin{document}

\title{The Naming Game in Social Networks: \\ Community Formation and Consensus Engineering%
\footnote{The original publication is available at \\
www.springerlink.com/content/70370l311m1u0ng3/}}

\titlerunning{The Naming Game in Social Networks}        

\author{\mbox{Qiming Lu \and G. Korniss \and B.K. Szymanski} }

\authorrunning{Q. Lu, G. Korniss, and B.K. Szymanski} 

\institute{Qiming Lu  \and G. Korniss (corresponding author) \at
Department of Physics, Applied Physics, and Astronomy,\\
Rensselaer Polytechnic Institute, 110 8th Street, Troy, NY 12180-3590, USA \\
\email{luq2@rpi.edu, korniss@rpi.edu} \and B.K. Szymanski  \at
Department of Computer Science and Center for Pervasive Computing
and Networking, \\
Rensselaer Polytechnic Institute, 110 8th Street, Troy, NY 12180-3590, USA \\
\email{szymab@rpi.edu} }

\date{}

\maketitle

\vspace*{-0.80truecm}
\begin{abstract}
We study the dynamics of the Naming Game [Baronchelli et al., (2006)
J. Stat. Mech.: Theory Exp. P06014] in empirical social networks.
This stylized agent-based model captures essential features of
agreement dynamics in a network of autonomous agents, corresponding
to the development of shared classification schemes in a network of
artificial agents or opinion spreading and social dynamics in social
networks. Our study focuses on the impact that communities in the
underlying social graphs have on the outcome of the agreement
process. We find that networks with strong community structure
hinder the system from reaching global agreement; the evolution of
the Naming Game in these networks maintains clusters of coexisting
opinions indefinitely. Further, we investigate agent-based network
strategies to facilitate convergence to global consensus.
\end{abstract}


\section{Introduction}

Agent-based models and simulations provide invaluable frameworks and
tools to gain insight into the collective behavior of social systems
\citep{Axtell1996,Challet2005,Anghel2004}. Opinion spreading and
social dynamics \citep{Durlauf,Castellano_RevModPhys} on regular and
random networks are examples of the latter. A large number of
studies have investigated models of opinion dynamics
\citep{Castellano2005,Eli,Deffuant,Krause_2002,Lorenz_2007,Lorenz_2008,KR2003,Sood,Sznajd,Kozma2007,Benczik2007,Antal2005,Jung2008}
and the dissemination of culture
\citep{Axelrod,SanMiguel_2005,Candia_IJMPC2007}, while fundamental
models for residential and ethnic segregation have also attracted
strong interest
\citep{Schelling_1971,Zhang_2004,Vinkovic_2006,Lim_2007}. Most
recently, researchers have also turned their focus to models where
both the network topology and opinions change over time
\citep{Kozma2007,Kozma2008_PRL}. With the availability of empirical
data sets and cheap and efficient computing resources, one can
implement stylized socio-economic models on empirical social
networks, and evolve ``artificial societies" \citep{Axtell1996} to
study the collective properties of these systems.

Here, we focus on one such stylized model, the Naming Game (NG)
\citep{Baronchelli_2005a}. The NG is a minimal model, employing
local communications, that can capture generic and essential
features of an agreement process in networked agent-based systems.
For example, in the context of a group of robots (the original
application), the NG dynamics mimics the emergence of shared
communication schemes (synthetic languages), while in the context of
sensor networks, such an agreement process can describe to the
emergence of a shared key for encrypted communications. In a system
of human agents, the NG can be considered as a minimal model to
describe the recent phenomenon of collaborative tagging or social
bookmarking \citep{Cattuto_2007a,Cattuto_2007b,Golder_2006} on
popular web portals like like Del.icio.us (http://del.icio.us),
Flickr (www.flickr.com), CiteULike (www.citeulike.org), and Connotea
(www.connotea.org). Another common example is the evolution and
spread of coexisting dialects in everyday use
(see, e.g., the geographical distribution of ``Pop" vs ``Soda"
for soft drinks in the US \citep{pop_vs_soda}). In
a broader context, the NG can be employed to investigate the emergence of
large-scale population-level patterns arising from empirically
supported local interaction rules between individuals.

The common feature in the above examples and applications is that
global agreement can emerge spontaneously (without global
enforcement) purely as a result of local (e.g., pairwise agent-to
agent) communications. The NG has been studied intensively on
regular and random complex network models (see next Section). Here
we investigate the evolution of the agreement process in the NG on
empirical social graphs. It is well known that empirical social
graphs exhibit strong community structure
\citep{Girvan_PNAS2002,Newman_PRE2004,Onnela_PNAS2007,Vicsek_Nature2005,Palla_Nature2007}.
It is also known that in networks with community structure, reaching global agreement can be
hindered \citep{Lambiotte_JSM2007,Candia_JSM2008}. Here, we
investigate the NG precisely from this viewpoint. Further, we
analyze strategies to destabilize otherwise indefinitely coexisting
clusters of opinions, to reach global consensus of a selected
opinion. The later can also be considered as an abstract agent-based
marketing approach.

The paper is organized as follows. In Sec.~2 we briefly review
recent results on the NG on various regular and complex networks
models. In Sec.~3, we present results for the NG on empirical social
networks. In particular, we investigate the effect of communities in
the underlying static social graphs on the agreement process (typically
leading to indefinitely coexisting clusters of opinions). In Sec.~4,
we study and analyze node-selection strategies to facilitate the
convergence to a global opinion. In Sec.~5, we conclude our paper
with a brief summary.

\section{Background, Model, and Prior Results on the Naming Game on Regular and Complex Network Models}

\begin{figure} [t]
\vspace*{3.5truecm}
\includegraphics{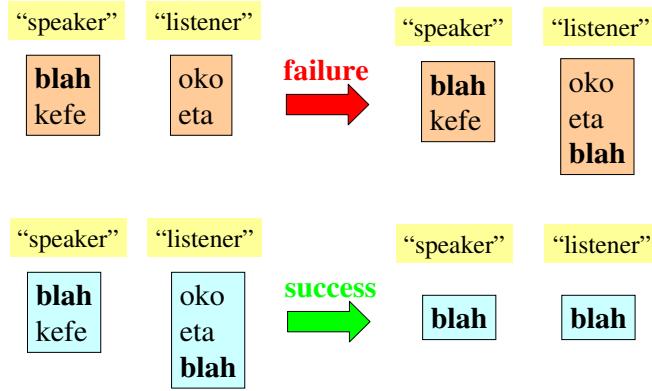}
\vspace*{2.0truecm}
\caption{Schematic rules of the Naming Game
\citep{Baronchelli_2005a} as described in the text. If the speaker
has more than one word on her list, she randomly chooses one; if it
has none, it generates one randomly.}
\label{NG_rules}
\end{figure}

In the simplified version of the NG, agents perform {\em pairwise}
games in order to reach agreement on the name to assign to a {\em
single} object. This version of the NG was investigated on the
complete graph (CG) (corresponding to mean-field or homogeneous
mixing) \citep{Baronchelli_2005a,Baronchelli_2005b}, on regular
\citep{Baronchelli_2006a}, on small-world (SW)
\citep{Baronchelli_2006b,Lin2006}, and on scale-free (SF) networks
\citep{Baronchelli_2006c,Baronchelli_2006d}. On a CG, each agent has
a chance to meet with all others and compare their current local
vocabularies (list of ``synonyms") before updating them. On regular
networks, agents have only a small number of nearest neighbors with
whom they can interact/communicate (e.g., four nearest neighbors in
two dimensions). The communication in both cases is ``local", in
that {\em pairs of agents} are selected to interact and to update
their vocabularies. The basic algorithmic rules of the NG are as
follows \citep{Baronchelli_2005a,Baronchelli_2006a}. A pair of
neighboring nodes (as defined by the underlying communication
topology), a ``speaker" and a ``listener", are chosen at random.\/%
\footnote{Note that on strongly heterogeneous (scale-free) graphs,
the order whether the listener or the speaker is chosen first,
strongly impacts the efficiency toward global agreement. Choosing
the listener first at random will increase the chance for selecting
a node (as a neighbor) with larger degree for speaker. In turn, hubs
will be the most frequent speakers, giving rise to faster
convergence to global agreement at a mildly elevated memory cost
\citep{Baronchelli_2006c,Baronchelli_2006d}.}
The speaker will transmit a word from her list of synonyms to the
listener. (If the speaker has more than one word on her list, she
randomly chooses one; if it has none, it generates one randomly.) If
the listener has this word, the communication is termed
``successful", and both players delete all other words, i.e.,
collapse their list of synonyms to this one word. If the listener
does not have the word transmitted by the speaker (termed
``unsuccessful" communication), she adds it to her list of synonyms
without any deletion. In this paper, we measure time in units during which
$N$ agents are selected at random as speakers, where $N$ is the number of agents in the network.
The above rules are summarized in Fig.~\ref{NG_rules}.

Here, we considered initial conditions when all agents have an empty
vocabulary. Then such an agent, when chosen to be a speaker, invents
a {\em random} word to be transmitted to the listener. (For initial
conditions with a single random word per agent, see the Electronic
Supplementary Material.)
In terms of the number of different words, the evolution of the game
will go through stages of growth (due to unsuccessful
communications) and stages of elimination (due to successful ones).
In all of the above mentioned networks, starting from empty
vocabularies, an early time explosion of words is followed by a slow
elimination of all synonyms, except one; that is agents come to
global agreement on the naming of the object in question.

Also recently, the NG was studied on homogeneous random geometric
graphs (RGGs) \citep{Lu_AAAI2006,LKS_PRE08}. RGG is both spatial and
random \citep{percolation,penrose,Dall_2002}. Motivated by the
deployment of sensor networks, nodes are randomly scattered in a
two-dimensional area, and two nodes are connected if they fall
within each others transmission range. Further, pairwise
communications were replaced by local broadcasts to capture the
essential features of communication protocols in sensor networks.
Similar to earlier findings on regular, SW, and SF networks, we
found that the NG on RGG with homogeneous node density also leads to
global consensus, facilitating an application to autonomous key
creation for encrypted communication in a community of sensor nodes
\citep{Lu_AAAI2006,LKS_PRE08}.

It was found that the NG dynamics on the above networks will lead to
global consensus among all agents, i.e., after some time, agents'
vocabularies eventually converge to a unique word, the same for all
agents
\citep{Baronchelli_2005a,Baronchelli_2005b,Baronchelli_2006a,Baronchelli_2006b}.
The major differences between the NG on CGs (homogeneous mixing)
and on low-dimensional networks (such as regular one- or
two-dimensional grids, and RGGs) arise in the scaling of the memory need
and in the scaling of the time $t_c$ with the number of agents $N$ to reach global agreement.
[The memory need in the present context is the typical value of the
largest number of words an agent may posses throughout the evolution
of the game \citep{Baronchelli_2005a,Baronchelli_2006a}.]
In CGs, the convergence process to global agreement is fast,
$t_c\sim{\cal O}(N^{1/2})$ (measured in units of
communications per agent), but large memory, ${\cal O}(N^{1/2})$, is
needed per agent \citep{Baronchelli_2005a}. For a regular
two-dimensional network or RGG, spontaneous evolution toward a
shared dictionary is slow, $t_c\sim{\cal O}(N)$, but the memory
requirement per agent is much less severe, ${\cal O}(1)$
\citep{Baronchelli_2006a}. When the NG is implemented on
Watts-Strogatz \citep{Watts98} SW networks
\citep{Baronchelli_2006b}, or when long-range random links are added
to the RGG \citep{LKS_PRE08}, the agreement dynamics performs
optimally in the sense that the memory needed is small, while the
convergence process is much faster than on the respective
low-dimensional network, $t_c\sim{\cal O}(N^{0.4})$, closer to that
of CGs or homogeneous mixing.

The above results, on spatial graphs, can be understood within the
framework of coarsening, a well know phenomenon from the theory of
domain and phase ordering in physical and chemical systems
\citep{Bray}. Starting from empty vocabularies, agents invent words
randomly. After time of ${\cal O}(1)$ [on average one communication
per node], ${\cal O}(N)$ {\em different} words have been created.
Following the early-time increase of the number of different words
(essentially corresponding to the number of different clusters of
agents) $N_d(t)$, through pairwise or local communications, agents
slowly reconcile their ``differences", and eventually will all share
the same word. First, a large number of small spatial clusters
sharing the same word develop. By virtue of the slow coalescence of
the interfaces separating the clusters, more and more of the small
clusters are being eliminated, giving rise to the emergence of
larger clusters, eventually leading to one cluster in which all
nodes are sharing the same word, i.e., $N_d$$=$$1$
\citep{Baronchelli_2006a,LKS_PRE08}. In domain coarsening, the
typical size of domains (each with already agreed upon one word) is
governed by a single length scale $\xi(t)\sim t^{\gamma}$ with
$\gamma$$=$$1/2$. Thus, in $d$ dimensions the average domain size
(inside which all agent share the same word) scales as
$\xi^d(t) \sim t^{d\gamma} $
and the total number of {\em different} words $N_d$ at time $t$
scales as the typical number of domains
$N_d(t) \sim N/\xi^d(t)\sim N t^{-d\gamma}$.
Global consensus is reached when $\xi^d(t_c)\sim N$ (or equivalently
$N_d(t)\sim 1$), hence the typical time to global agreement scales
as $t_c\sim N^{1/(d\gamma)}$.

On SW and SF random network models with no community structure, the
long-time behavior of the NG is essentially governed by the
mean-field fixed point,\/%
\footnote{Here, by the mean-field fixed point, we refer to the
characteristic scaling behaviors of the NG on the complete graph
(also referred to as homogenous mixing) where each agent can
potentially interact with all others
\citep{Baronchelli_2005a,Baronchelli_2006c}.}
and global consensus time scales as $t_c\sim N^{1/2}$ (although with
noticeable finite-size corrections)
\citep{Baronchelli_2006b,Baronchelli_2006c,LKS_PRE08}. On the other
hand, \citet{Baronchelli_2006c} found that on stylized network
models with community structure (composed of fully connected cliques
with a single link between cliques) the evolution of the NG runs
into long-living meta-stable configurations, corresponding to
different co-existing words (different for each clique). Here we
study precisely this later scenario by implementing the NG on static
empirical social graphs.

In this work we employ the basic NG where agents have infinite
memories (i.e., the number of words they can store on their
individual lists is unlimited). For a finite-memory version
\citep{Wang_2007} of the model on social networks see the Electronic
Supplementary Material.

In passing we note that the issue of the emergence of meta-stable or
frozen opinion clusters and fostering consensus have been discussed
for models of opinion formation under bounded confidence
\citep{Deffuant,Krause_2002,Lorenz_2007,Lorenz_2008}. In those
models, however, community formation or opinion segmentation is the
result of the agents' interaction being limited by bounded
confidence: an agent can gradually adjust her opinion toward another
one's only if their opinions were already sufficiently close to one
another to begin with. As a result, opinion segmentation can emerge
in networks with no community structure with low-confidence agents.
In contrast, the NG dynamics does not require that agents' opinions
are sufficiently close in order to potentially interact (i.e., their
confidence is unbounded), and as mentioned earlier, the NG dynamics
always lead to global consensus on networks {\em without} community
structure. Our motivation here, by studying the NG on empirical
social graphs, is to directly study how the community structure of
the underlying graphs affects the emergence of meta-stable or
long-living opinion clusters.

\section{The Naming Game on Empirical Social Networks}

One of the most important feature of social graphs is their
modularity: these networks typically consist of a number of
communities; nodes within communities are more densely connected,
while links bridging communities are sparse. Since the community
structure of empirical networks is often not known a priori, detecting
communities in large networks itself is a difficult problem \citep{Vicsek_Nature2005}.
A number of current methods for finding community structures utilize
various forms of hierarchical clustering, spectral bisection methods
\citep{Scott_2000,Newman_PRE2006,Huberman_EPJB2004}, and
iterative high-betweenness edge removal \citep{Newman_PRE2004,Newman_EPJB2004,Girvan_PNAS2002}.
A different approach involves searching for the ground-states of generalized multi-state
spin models (corresponding to different opinions) on these networks,
such as the $q$-state Potts model \citep{Domany1996,Bornholdt2004,Kertesz2007,Barthelemy2007}.
Also, recently a novel method has been developed to detect overlapping
communities in complex networks \citep{Vicsek_Nature2005}.

The NG, as summarized in the Sec. 2, in low-dimensional networks exhibits
slow coarsening, while networks with small-world characteristic
(small shortest path, such as in SW and SF networks) facilitate
faster (and guaranteed) convergence to a global consensus among
nodes. But in all cases, global consensus is reached, provided the
network has no heterogeneous clustering or modularity (i.e.,
community structure).

Here, we study the NG on networks which do exhibit strong community
structure. The set of social networks (high-school friendship
networks), on which we implemented the NG, were provided by the
National Longitudinal Study of Adolescent Health (Add Health)%
\footnote{ This research uses the network-structure data sets from
Add Health, a program project designed by J. Richard Udry, Peter S.
Bearman, and Kathleen Mullan Harris, and funded by a grant
P01-HD31921 from the National Institute of Child Health and Human
Development, with cooperative funding from 17 other agencies. For
data files contact Add Health, Carolina Population Center, 123 W.
Franklin Street, Chapel Hill, NC 27516-2524, (addhealth@unc.edu,
http://www.cpc.unc.edu/projects/addhealth/. }.
The high-school friendship networks investigated here, were
constructed from the results of a paper-and-pencil questionnaire in
the AddHealth project \citep{Moody_AJS2001}. Here, nodes represent
students while the edges are for their mutual relations or
friendships. Two students are considered to be friends (thus have a
link between them) when one nominates the other as her/his friend
and both of them participated in some activities, e.g., talked over
the phone, spent the weekend together, etc., in the last seven days.
(for this study, we considered the relationships reciprocal, and
associated them with undirected links in the NG).
These networks exhibit exponential degree distributions (no hubs), with an average degree of the order of $10$.
For a baseline comparison we also constructed
a Watts-Strogatz (WS) network \citep{Watts98} network with the same
number of nodes $N$, average degree $\overline{k}$, and clustering coefficient
$\overline{C}$ as the friendship network. The WS network has
homogeneous clustering, hence, no community structure.
\begin{figure}[t]
\vspace*{3.0truecm}
\includegraphics{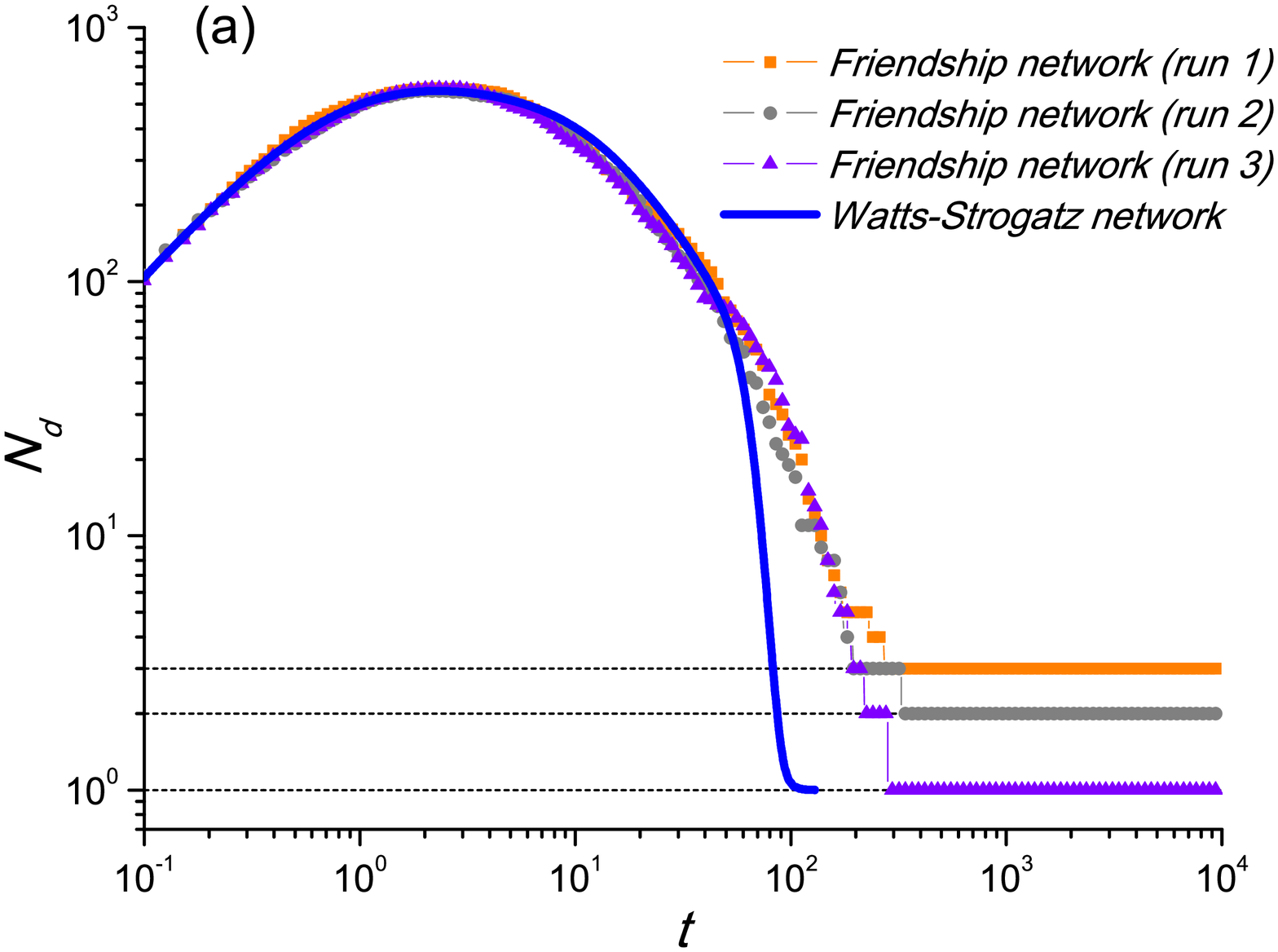}
\includegraphics{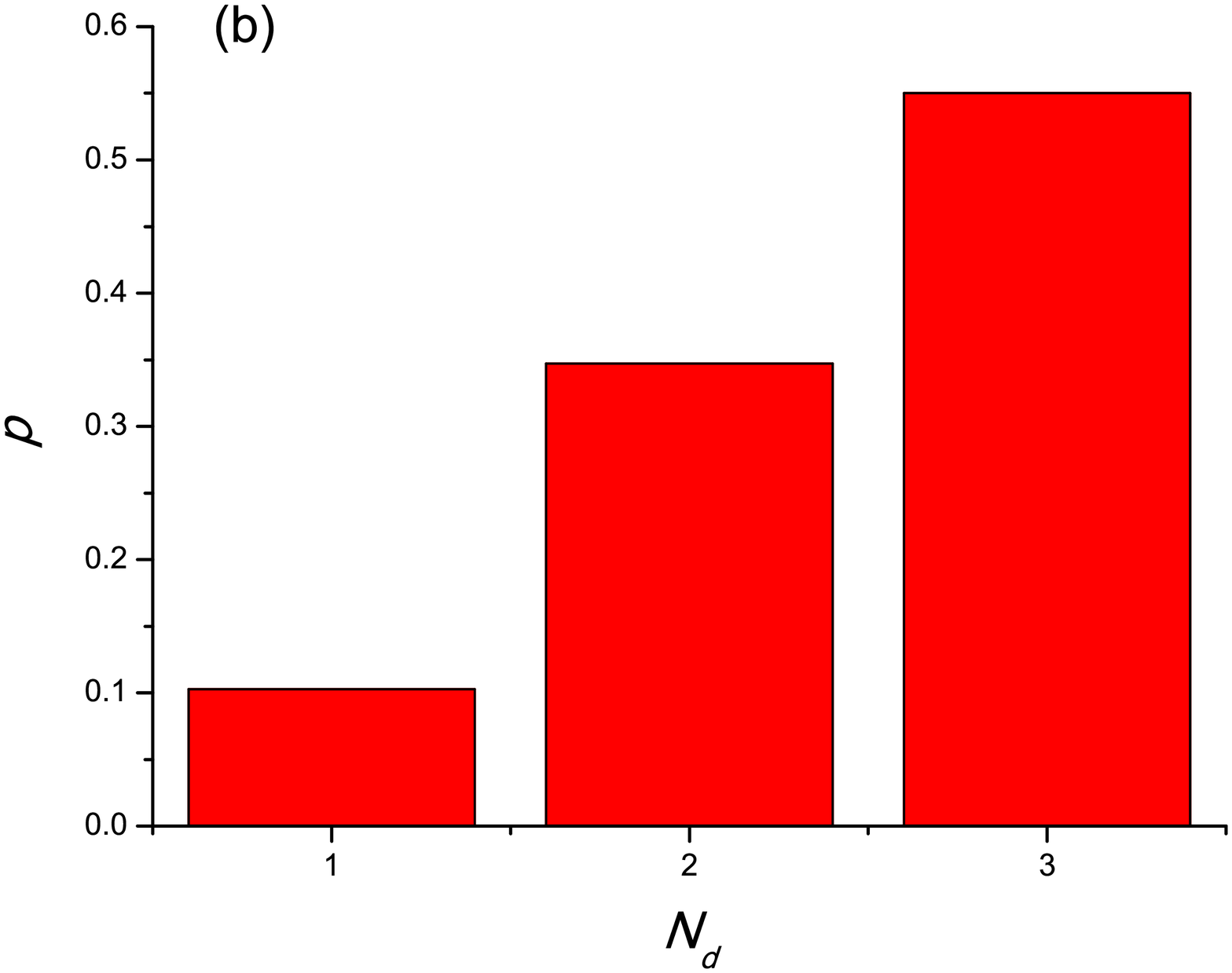}
\vspace*{3.5truecm}
\caption{(a) Number of different words $N_d$ vs
time for a friendship network (thin lines) and for the Watts-Strogatz network (bold line).
$N=1,127$, $\overline{k}=8.8$, and $\overline{C}=0.067$ for both systems.
Results for the WS network are averaged over $1000$ independent realizations.
For the high-school friendship network we show three individual realizations (thin lines),
reaching different final states with $N_d=1$, $N_d=2$,
and $N_d=3$ (indicated with horizontal dashed lines).
Note the log scales on both axes.
(b) The probability (relative frequency) of final configurations with $N_d$ different words (opinions)
for the same high school friendship network as in (a) based on $10,000$ independent runs.
Statistically, in this particular network, the most likely
final configuration exhibits three opinions.}
\label{fig.nd-HSN-WS}
\end{figure}

\begin{figure}[t]
\hspace*{0.0cm}
\includegraphics[width=4.6in]{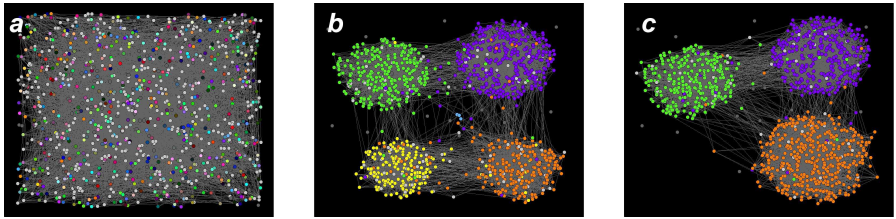}
\vspace*{-0.00truecm}
\caption{Snapshots of the time-evolution of the Naming Game on a high-school friendship network.
Initially agents have an empty word list (no opinions). In the snapshots, different colors correspond to different words.
In the very early stage of the game (a), ``speakers" with no words has to create one randomly.
After a slow but steady coarsening of opinions, in the final stages of the game, the system exhibit
relatively long plateaus in the number of different opinions.
The corresponding clusters, i.e., agents with the same opinions, can be regarded as communities.
For the particular network shown here, in the next to final stage (b),
the network exhibits four communities. Eventually,
two of these communities coalesce, leading to a final configuration (c) with three communities.}
\label{fig.HS_snapshots}
\end{figure}

We selected a few networks with a large number of students (on the
order of 1,000) from the available data set. Starting from an empty
word list for all agents, both the friendship network and WS network
show nearly identical early-time development of the number of
different words $N_d$. However, the friendship-network simulations
exhibit a long-time behavior very different from the ones discussed
in Section 2, and also from the baseline reference, the NG on the WS
network [Fig.~\ref{fig.nd-HSN-WS}(a)]. In the late stage of the NG,
networked agents without community structure (including the WS
network) {\em always} exhibit a spontaneous evolution toward a
shared ``dictionary" (or opinion), i.e., a global consensus is
reached. In contrast, in the empirical high-school networks,
consensus is rarely reached (for long but finite simulation times)
[Fig.~\ref{fig.nd-HSN-WS}(a)]. For this particular high-school
friendship network, performing $10,000$ independent runs of the NG
with a fixed simulation time of $t=10^4$ steps, $10\%$, $35\%$, and
$55\%$ of these runs, ended up with one, two, and three different
words, respectively, in their final configurations
[Fig.~\ref{fig.nd-HSN-WS}(b)]. Thus, in this network, the most
likely (or typical) outcome of the NG is one with three different
clusters of opinions. Snapshots taken from the typical evolution of
the NG on this network are shown in Fig.~\ref{fig.HS_snapshots}.
In analogy with domain formation in physical systems, we can regard
these long-living configurations with coexisting multiple opinions
as ``meta-stable" ones.

The emergence of different long-living clustered opinions is not
unexpected. In fact, the same high-school networks have been
analyzed for community structures in a study of friendship
segregation along racial lines among high-school students
\citep{Moody_AJS2001,Gonzales_2007}. For example, close to the final stage,
the time-evolution of the NG on the particular network shown in
Fig.~\ref{fig.HS_snapshots}(b) exhibits four communities.
These four clusters of opinions correspond to segregation along the
two-schools involved in the particular network, high-school (HS) --
middle-school (MS) pair, and along racial lines, whites students --
black students in each. Checking the race and school-grade attribute
of the node information in the raw data, we confirmed that the four
communities exhibited by the NG in Fig.~\ref{fig.HS_snapshots}(b)
correspond to black HS (upper left), white HS (upper right), black
MS (lower left), and white MS (lower right) students.
Then, in the final state [Fig.~\ref{fig.HS_snapshots}(c)], only
three communities remain; opinions, segregated along the racial line
coalesce in the middle-school portion of the students, simply
indicating that racial segregation in friendships is weaker in this
group, in this particular network set.

Admittedly, the objective of our paper is not to draw over-ambitious
conclusions from a social science viewpoint. Instead, we are
interested in how the evolution of the NG (a stylized model for
opinion formation) is affected by the community structure of the
underlying graphs, such as the high-school friendship networks which
are well-known to exhibit strong community structure
\citep{Moody_AJS2001,Gonzales_2007}. We demonstrated that the
outcome of the NG, is strongly affected by the existence of
communities in the underlying network. Conversely, at some coarse
level, the long-living late-stage meta-stable clusters of words
(opinions) reveal important aspects of the community structure of
the underlying network. Thus, the NG, together with other stylized
models for opinion formation, can not only be used as a tool to
understand generic features of spontaneous agreement processes in a
network of artificial or human agents, but can also be employed to
extract relevant information on the community structure of complex
networks
\citep{Domany1996,Bornholdt2004,Kertesz2007,Barthelemy2007}.

\section{Engineering Consensus in Social Networks}

There are several ways to influence the outcome of social dynamics,
e.g., to facilitate the outcome of a specific global opinion that
one would prefer the system to achieve (preferred opinion for
short). All methods essentially rely on ``breaking the symmetry" of
the otherwise equivalent coexisting opinions. One possibility is to
expose and couple many or all agents to an ``external" global signal
(analogous to mass media effects)
\citep{Candia_IJMPC2007,Candia_JSM2008}. Alternatively, one can
break the symmetry by choosing a small number of well-positioned
``committed" agents who will stick to a preferred opinion without
deviation. In the next subsection, we investigate this latter
scenario first.

\subsection{Committing Agents}
\begin{figure}[t]
\hspace*{0.0cm}
\includegraphics[width=4.6in]{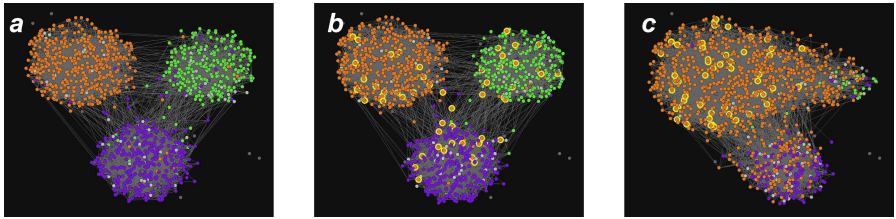}
\vspace*{-0.00truecm}
\caption{Snapshots of the Naming Game on a
high-school friendship network with committed agents. The system is
initialized from a state with three coexisting meta-stable
communities [see (a)] with a small number of well-positioned
committed agents (indicated with yellow core around the nodes as
indicated in (b). Global consensus (i.e., a single opinion) is
reached exponentially fast. Here we employed $50$ committed agents,
selected according to their degree ranking.}
\label{fig.HS_consensus}
\end{figure}

In the simulations, by committed agents we mean an agent who has a
fixed opinion which cannot be changed. In the context of the NG, a
committed agent has a single word. As a listener, she does not
accept any new word from their neighbors, but as a speaker, always
transmits her word. Of the three co-existing communities at the
end-stage of the NG [Fig.~\ref{fig.HS_snapshots}(c)], we choose one
community as the one representing the ``preferred" opinion, and we
``indoctrinate" selected committed agents with this opinion.
Fig.~\ref{fig.HS_consensus} shows snapshots of the evolution of the
NG with committed agents. Initiating the simulations from the final
configuration of the original NG (exhibiting three meta-stable
opinion clusters), introducing a small number of committed agents
yields a relatively fast convergence to the global consensus of the
selected opinion.

To quantify this phenomena we investigated the temporal behavior of
this agreement process, in particular, its dependence on the method
of selecting committed agents and on the number of these selected
agents. Among the methods to select committed agents are selecting
nodes with the highest degrees (nodes with the highest number of
neighbors), with the highest betweenness (likely to bridge different
communities), with hop-distance proximity to the core cluster (nodes outside, but no farther than two
hops from the core cluster of ``preferred" opinion), and for
comparison, also selecting committed agents at random.

Our main observation is that once the number of committed agents is
sufficient to induce global consensus, it happens {\em
exponentially} fast, independently of the selection method. More
precisely, we ran $10,000$ realizations of the NG with committed
agents. The initial configuration here is the final multi-opinion
meta-stable configuration of the original NG with no committed
agents (with $N_d=3$) [Fig.~\ref{fig.HS_consensus}]. We kept track
of the {\em fraction of surviving runs}, $n_s(t)$, defined as the
fraction of runs that have not reached global consensus by time $t$,
i.e., runs that have more than one opinion at time $t$. (This
quantity then can also be interpreted as the probability that a
single run has not reached consensus by time $t$.)
\begin{figure} [t]
\vspace*{3.0truecm}
\includegraphics{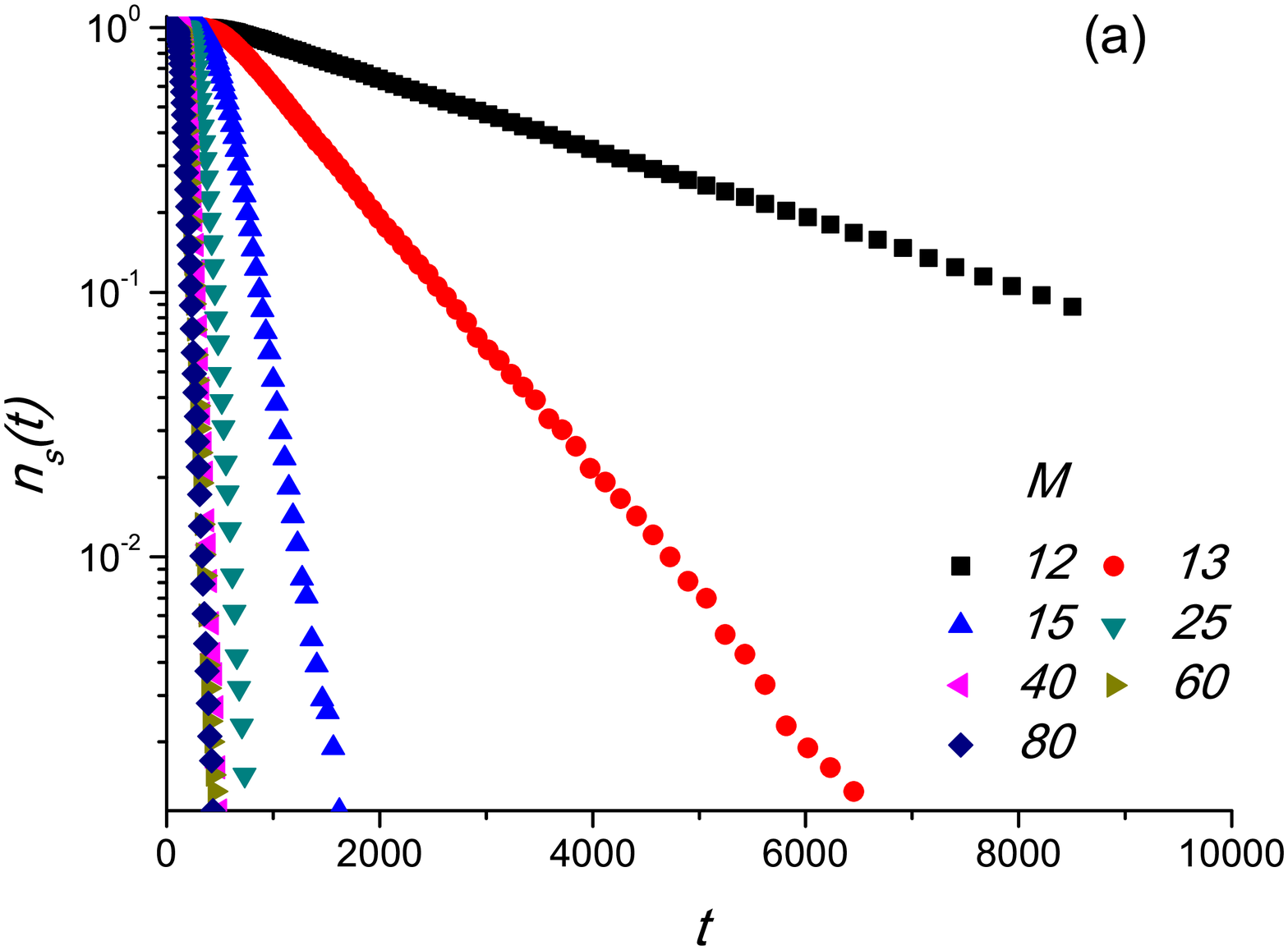}
\includegraphics{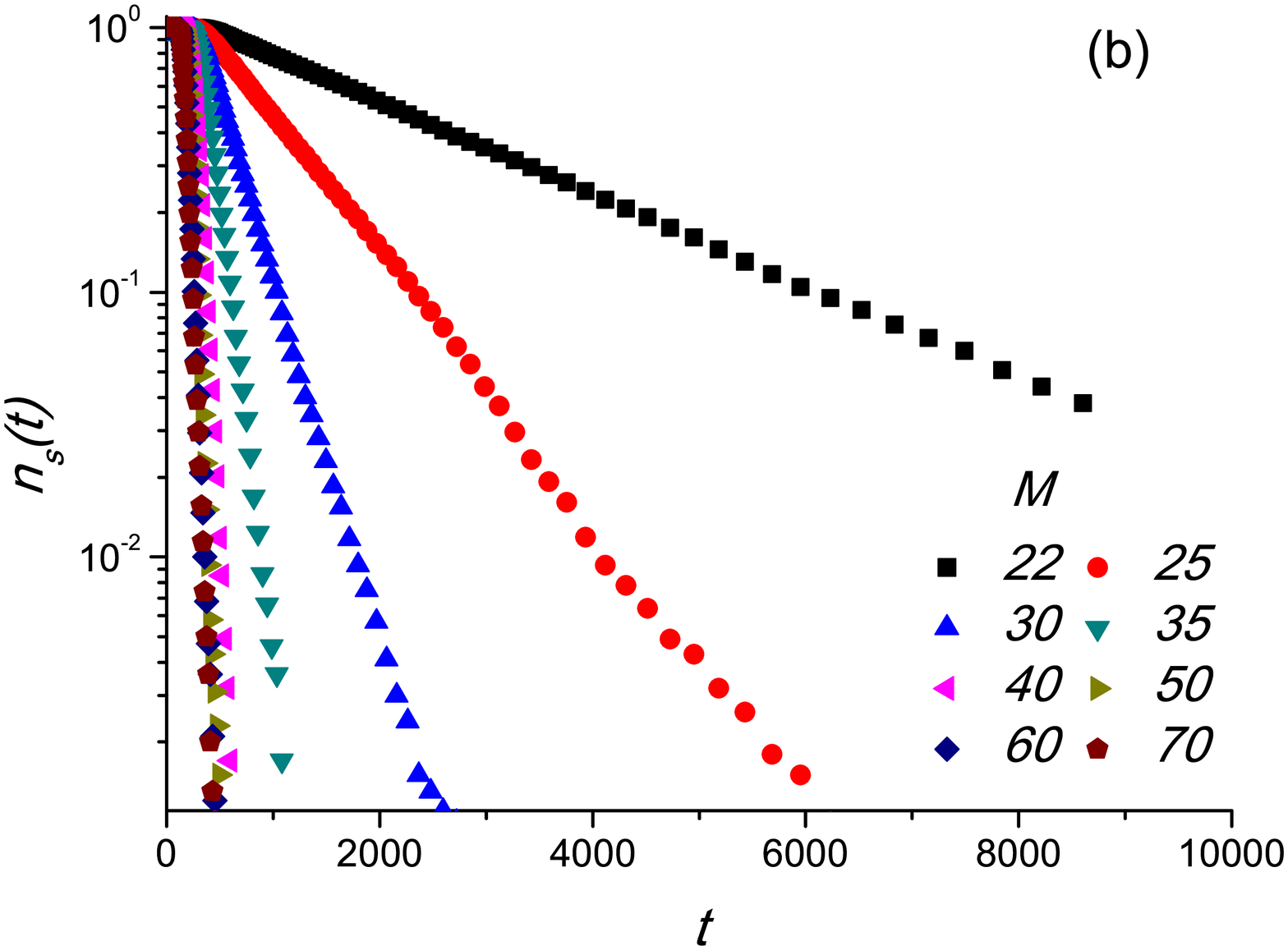}
\vspace*{3.5truecm}
\caption{Fraction of surviving runs as a function of time for varying number of committed agents $M$
when agents are selected according to their
(a) degree ranking and
(b) (shortest-path) betweenness ranking. The total number of agents is $N=1,127$.
For the  degree-based ranking selection method different symbols represent the fraction of surviving runs
for $12$, $13$, $15$, $25$, $40$, $60$, and $80$ committed agents, from top to bottom.
In betweenness selection method the number of committed agents $M$ ranges from
$22$, $25$, $30$, $35$, $40$, $50$, $60$, to $70$, from top to bottom.}
\label{fig.HS_degree_bcsp}
\end{figure}
We choose committed agents, to maximize their influence in reaching
global consensus, according to their ranking in a number of graph
theoretical measures. We selected the top $M$ agents according to
their degree, shortest-path betweenness centrality,
\citep{Newman_PRE2004,Newman_EPJB2004}, hop-distance proximity to the
preferred core opinion cluster, or at random, for reference.
Figure~\ref{fig.HS_degree_bcsp} displays the fraction of surviving
runs $n_s(t)$ for the degree and for the betweenness ranking for a
number of different committed agents.

A common feature of all methods is that a very small fraction
($f=M/N$) of committed nodes is sufficient to induce global
consensus. I.e., there seems to be a very low threshold in $f_{c1}$,
such that for $f>f_{c1}$ the dynamics with committed nodes leads to
global agreement. Further, in this case, the fraction of surviving
runs (fraction of runs with more than one opinion), $n_s(t)$, in the long-time regime, decays exponentially
\begin{equation}
n_s(t)\propto e^{-t/\tau} \;.
\label{n_s}
\end{equation}
The time scale of the exponential decay $\tau$, of course, depends
on the selection method and the fraction of committed nodes. The
inverse time scale $1/\tau$, i.e., the rate at which global
consensus is approached is, initially, an increasing function of the
number of committed nodes, but it quickly {\em saturates} and
essentially remains constant. This can be seen in
Fig.~\ref{fig.HS_degree_bcsp}, as the slopes of the exponential
decays are becoming progressively steeper, up to a certain $M$, then
they remain constant. Thus, there is second characteristic fraction
of committed agents, such that for $f>f_{c2}$ the rate of reaching
global consensus becomes essentially a constant (saturates).
\begin{figure}[t]
\hspace*{-0.5cm}
\includegraphics[width=3.7in]{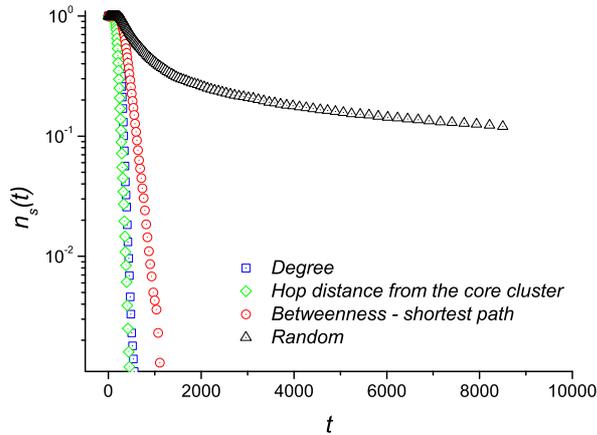}
\vspace*{0.00truecm} \caption{Fraction of surviving runs as a
function of time for different strategies with the same number of
committed agents on the same network ($M=35$, $N=1,127$,
$f\simeq0.031$.). The three strategies (selection of committed
agents) shown here are based on degree ranking (squares),
hop-distance proximity to the core cluster (diamonds), and
shortest-path betweenness (circles). For comparison, the result of
selecting committed agents randomly is also shown (triangles).}
\label{fig.hs_compare}
\end{figure}

\begin{figure}[t]
\hspace*{-0.5cm}
\includegraphics[width=3.7in]{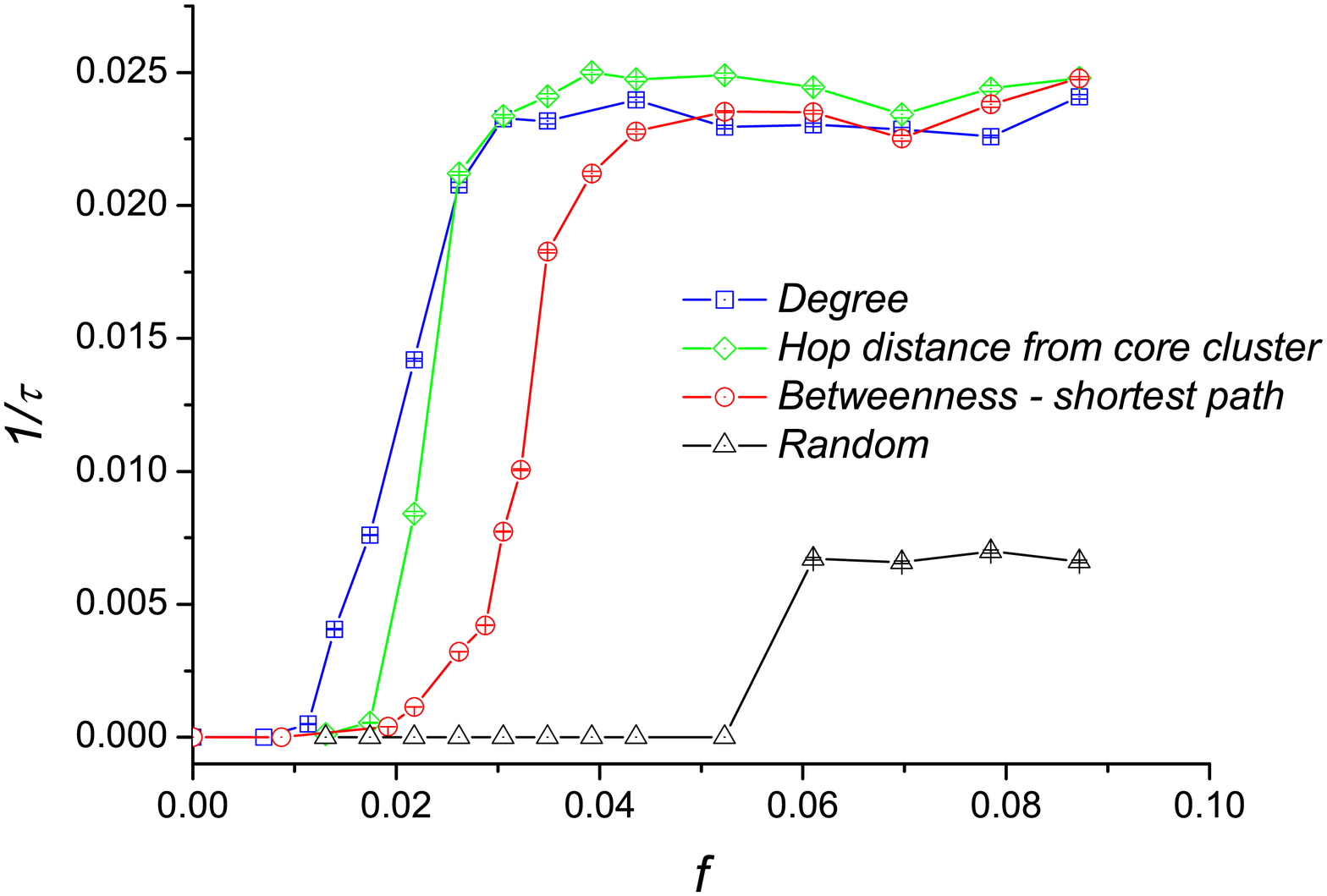}
\vspace*{0.00truecm}
\caption{Convergence rate as a function of the fraction of committed agents $f$$=$$M/N$,
for different selection methods of committed agents,
including the degree ranking (squares), hop-distance proximity to core cluster (diamonds),
and shortest-path betweenness (circles).
For comparison, the result of selecting committed agents randomly is also shown
(triangles).}
\label{fig.hs_slope}
\end{figure}

These three features, ({\it i}) small threshold $f_{c1}$ required
for global consensus, ({\it ii}) exponential decay of $n_s(t)$ if
$f>f_{c1}$ [Figs.~\ref{fig.HS_degree_bcsp} and
\ref{fig.hs_compare}], and ({\it iii}) saturation of the rate to
reach consensus for $f>f_{c2}$ [Fig.~\ref{fig.hs_slope}], are
exhibited by all selection method we considered here. Further, both
characteristic values and the gap between them are very small,
$f_{c1}$, $f_{c2}$, $f_{c2}$$-$$f_{c1}$$ \ll1$. These results are
essentially summarized in Fig.~\ref{fig.hs_slope}. The convergence
rate for the randomly selected committed nodes is also shown for
comparison. On this particular social network, selecting a small
number of the nodes with the highest degree works best, followed by
the hop-distance proximity (to the core cluster) ranking. (We refer
to a strategy as more efficient if the convergence rate $1/\tau$ is
larger for the same fraction of committed agents.) For example,
selecting the committed agents according to their degree ranking,
$f_{c1}\approx 0.01$ and $f_{c2}\approx0.03$
[Fig.~\ref{fig.hs_slope}]. Selecting committed agents just above
this latter fraction is optimal, since the rate of convergence does
not improve beyond this value.

In general, the optimal selection method will vary, depending on the
community structure of the particular underlying network. However,
because we changed the dynamics of the NG by breaking the symmetry
of otherwise equivalent opinions, the exponential decay and the
saturation of the convergence rate is expected to be generic for a
large class of opinion formation models on networks with community
structure.

\subsection{Global External Influence}

\begin{figure}[t]
\vspace*{3.0truecm}
\includegraphics{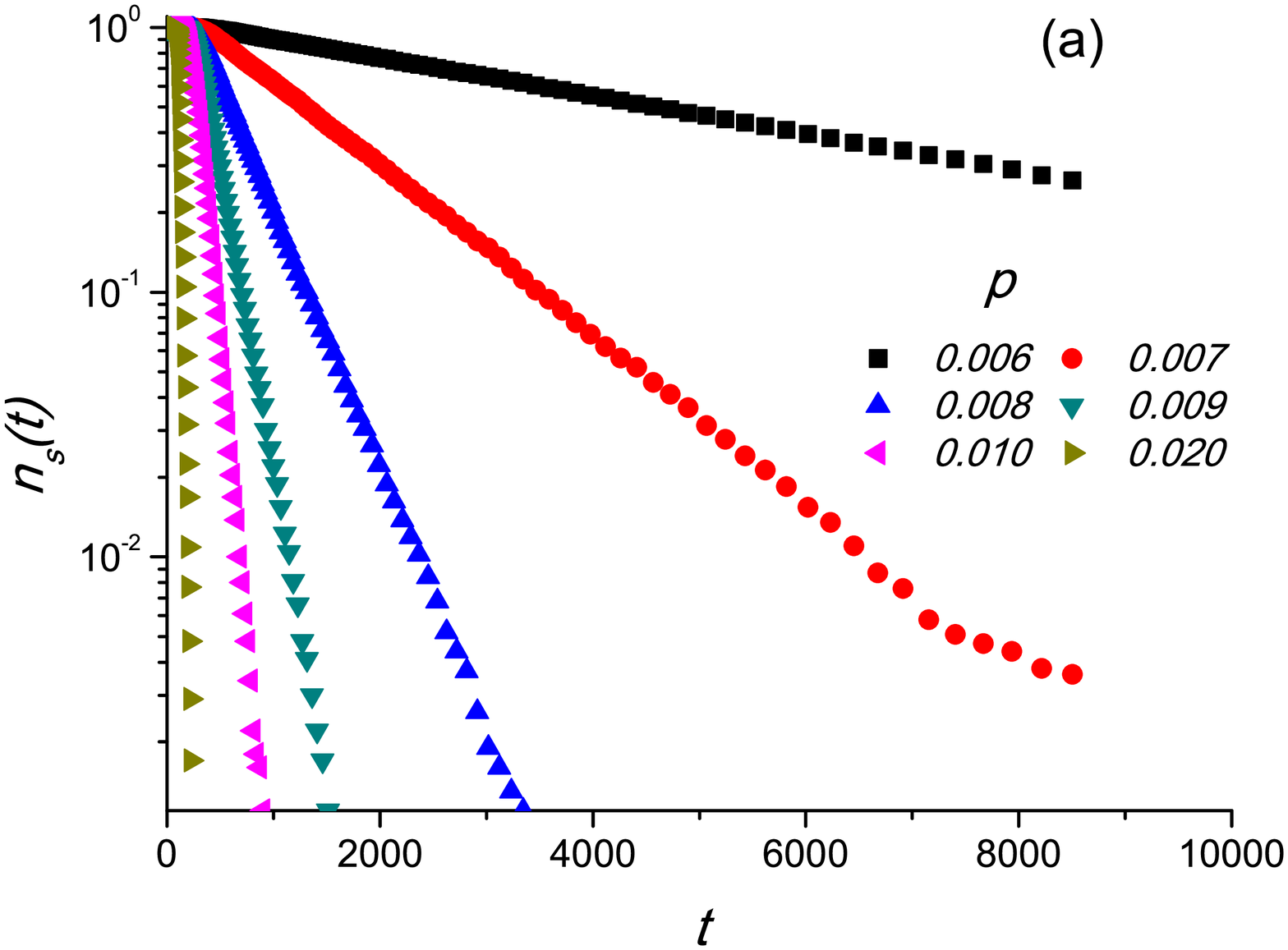}
\includegraphics{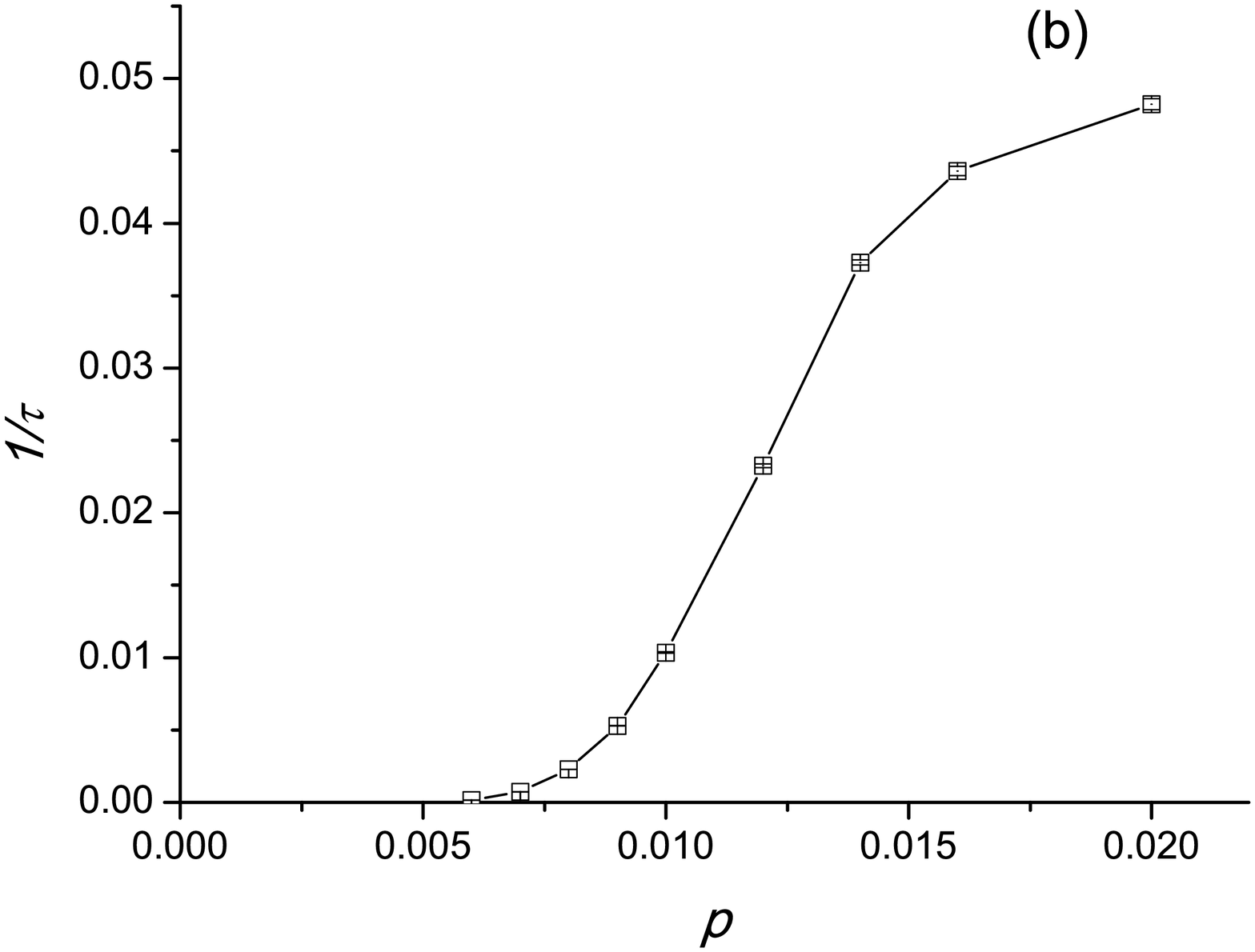}
\vspace*{3.5truecm}
\caption{
(a) The fraction of surviving runs as a function of time for several values of the strength of external influence $p$
($p$ is the probability that in a time step an agent will adopt the fixed externally and globally promoted opinion).
(b) Convergence rate to global consensus as a function of the strength of external field $p$. }
\label{fig.HS_external}
\end{figure}

As mentioned in the introductory paragraph of this Section, another
natural way of influencing the outcome of the competition among
otherwise neutral and meta-stable opinions, is to couple all or a
fraction of agents to a global external ``signal" [mimicking a mass
media effect \citep{Candia_IJMPC2007,Candia_JSM2008}]. For
comparison, we implemented the NG with an external field (affecting
all agents) corresponding to the selected opinion among the three
meta-stable ones in the final stage of the NG. Then, similar to the
committed-agent approach, we initialize the system with that final
meta-stable state with co-existing opinions of the original NG. In
the presence of mass media, an agent, when randomly chosen, with
probability $p$ will adopt the externally promoted opinion.
Otherwise, the usual rules of the game are invoked (i.e., the node,
as a speaker, initiates communication with a listener).
Our findings indicate that even for extremely small values of $p$,
the fraction of surviving runs (the fraction of runs that have not
reached global consensus) decays exponentially, ultimately leading
to global order [Fig.~\ref{fig.HS_external}(a)]. The rate of
convergence $1/\tau$ increases monotonically and smoothly with $p$
[Fig.~\ref{fig.HS_external}(b)].
For application oriented studies, one should
associate a cost with the mass-media coupling, and a cost with
committing an agent (e.g., finding these nodes and giving them
incentives impossible to resists), then perform a relevant
cost-benefit comparative analysis for the selection or optimal combination of two approaches.

\section{Summary}

We studied the Naming Game on social networks. Earlier works have
shown that this simple model for agreement dynamics and opinion
formation always leads to global consensus on graphs with no
community structure. On the other hand, social networks are known to
have rich community structure. The Naming Game on such networks
exhibits, in the late-stage of the dynamics, several meta-stable
coexisting communities; these configurations, in effect, are the
computationally observed final configurations.

In the context of models for social dynamics, communities manifest
themselves in the context in which distinct stylized opinions (e.g.,
religions, cultures, languages) have evolved and emerged over time.
Clusters of nodes having reached consensus are part of a community,
reflecting the inherent community structure of the underlying social
networks. Thus, if at the late stages of the social dynamics on the
networks, several communities persist (different opinions survive),
they are the authentic signatures of the underlying community
structure. The Naming Game, together with other similar models for
opinion formation, can be employed to probe these properties of
complex networks.

We then considered the task of destabilizing the coexisting
meta-stable opinions (in order to reach consensus) by selecting the
optimal number of committed agents with a preferred opinion, as an
alternative to a global external signal (mass media effect). The
results implied that a small number of committed agent is sufficient
to facilitate an exponential decay toward global consensus of the
selected opinion. Further, selecting more agent than a
system-specific upper cut-off, yields no improvement in the
convergence rate. Hence, there seems to be an optimal number of
agents for this task, beyond which it does not pay off to invest in
committing more agents. Selecting the committed agents according to
their degree, betweenness, or hop-distance proximity to the core cluster of
the preferred opinion, all displayed the above qualitative features.
Further, they all significantly outperformed committing the same
number of agents at random.

\section*{Acknowledgments}
We thank Y.-Y. Ahn, E. Anshelevich, and A.-L. Barab\'asi for
discussions. This work was supported in part by the  Office of Naval Research Grant No. N00014-09-1-0607.
The content of this paper does not necessarily reflect the position or policy of the U.S. Government,
no official endorsement should be inferred or implied.




\newpage

\renewcommand{\figurename}{Fig. S\hspace*{-0.10truecm}}
\setcounter{figure}{0}
\setcounter{page}{1}

\section*{Electronic Supplementary Material}

\vspace*{0.50truecm}
\section*{The Naming Game with one-word-per-agent initial conditions}

\begin{figure}[htbp]
\vspace*{3.0truecm}
  \includegraphics{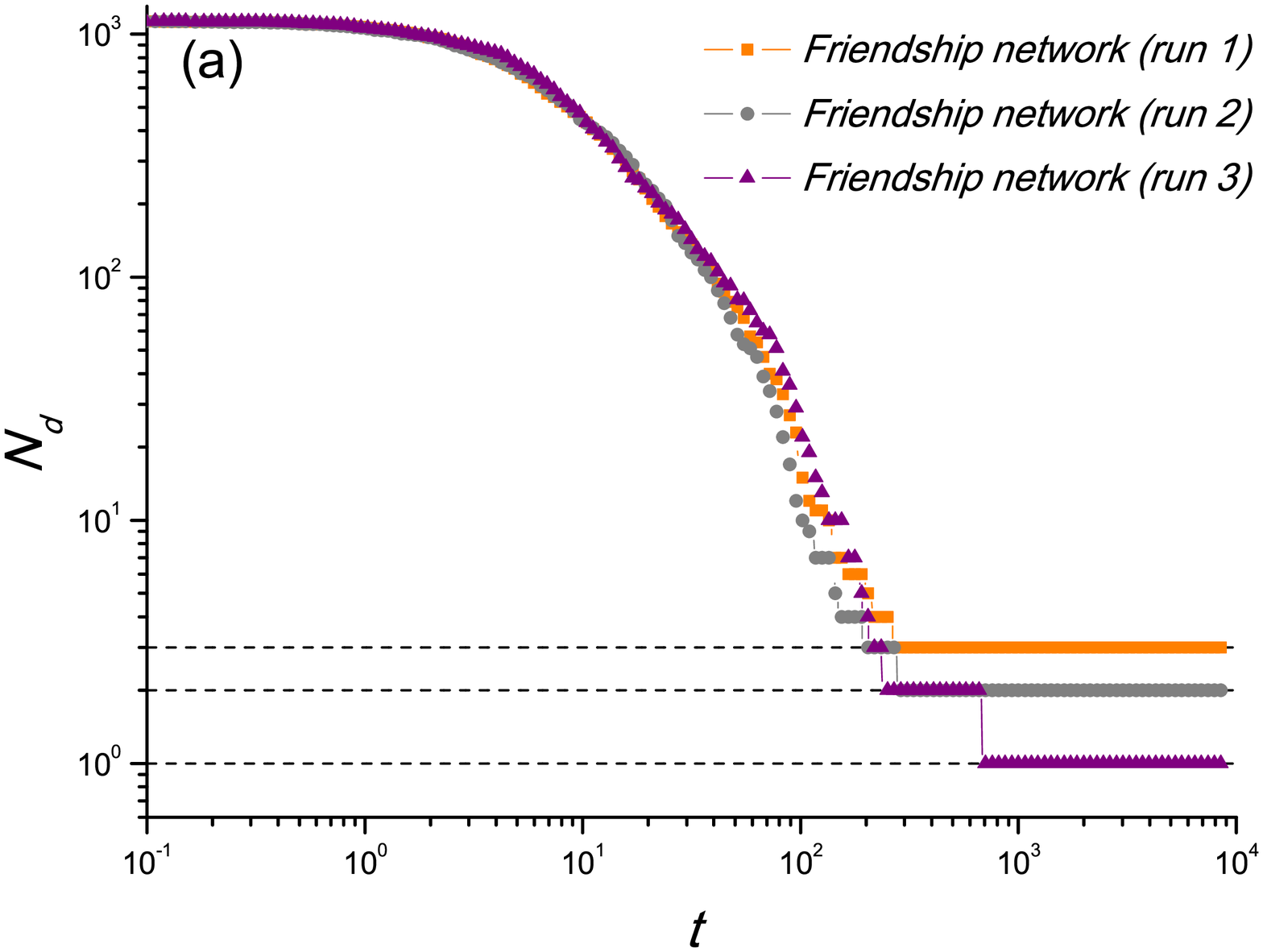}
  \includegraphics{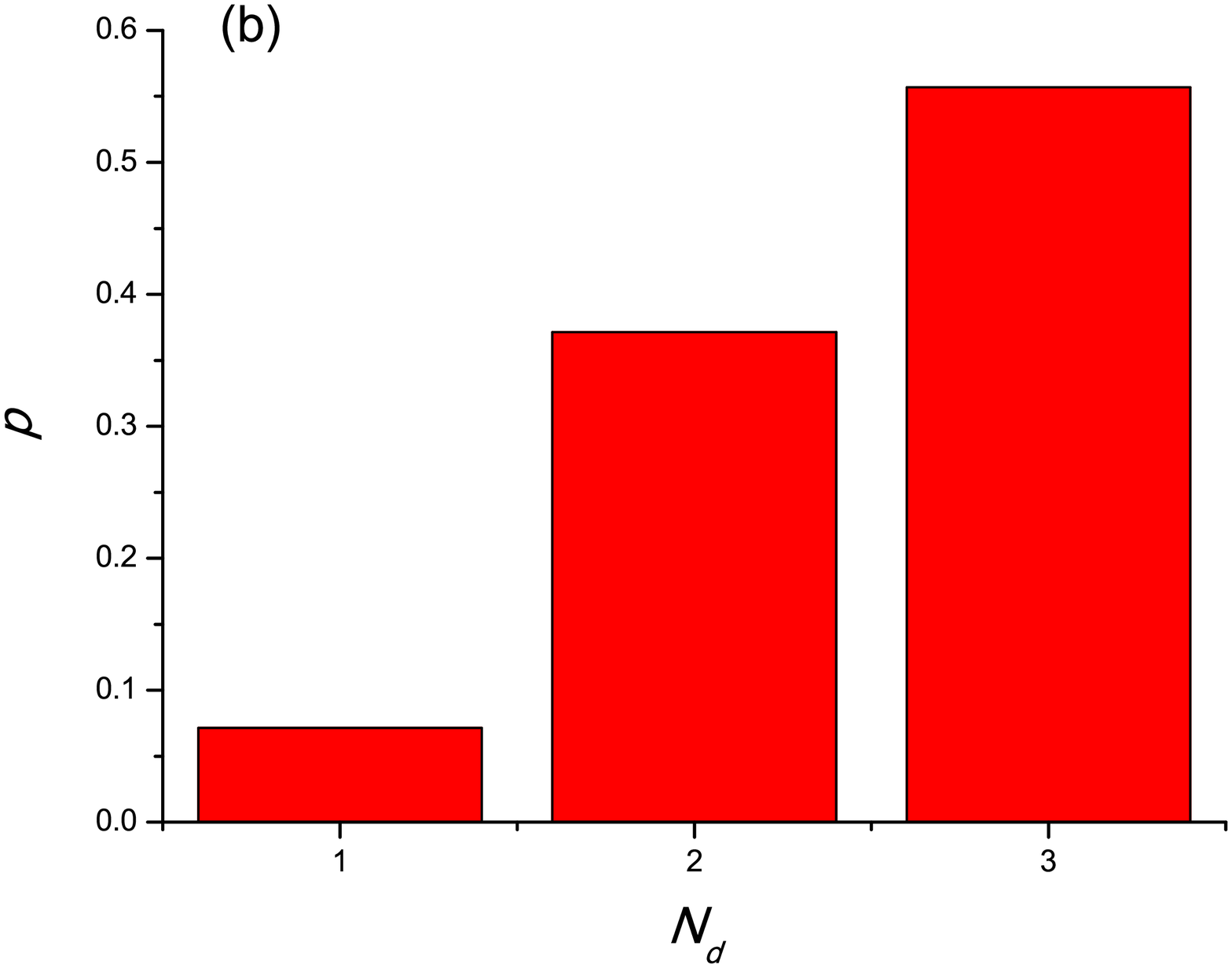}
\vspace*{3.5truecm}
\caption{The Naming Game on a friendship network where the
simulations are initialized from configurations with a single word
per agent (different for each agent). (a) Number of different words
$N_d$ vs time for a friendship network with $N=1,127$,
$\overline{k}=8.8$, and $\overline{C}=0.067$ (same friendship
network as in Fig.~2 of the main article). Results are shown for
three individual realizations, reaching different final states with
$N_d=1$, $N_d=2$, and $N_d=3$ (indicated with horizontal dashed
lines). (b) The probability (relative frequency) of final
configurations with $N_d$ different words (opinions) for the same
friendship network as in (a) based on $10,000$ independent runs.}
\label{fig_S1}
\end{figure}

%
Here, we show the behavior of the Naming Game (NG) with initial
configurations where each agent has exactly one word (or opinion),
different for each agent ($N_{d}(0)=N$). The rest of the rules of
the NG are the same as described in Sec.~2 of the main article.
Since each agent has a word initially, no new words will be
invented, and there is no growth phase ($N_{d}$ is monotonically
decreasing as a function of time); pairwise communications will lead
to the gradual elimination of existing opinions. At the end, only a
few opinion clusters remain, again, reflecting the community
structure of the underlying graphs. There is no significant
difference in the {\em late-stage scaling behavior} between the
empty-dictionary and the one-word-per-agent initial conditions: In
the former case, it takes of ${\cal O}(1)$ time steps to reach the
maximum of $N_{d}$, $N_{d}^{\rm max}\sim N/2$ [Fig.~2(a) of the main
article], after which slow ``opinion coarsening" begins. In the
latter case, the number of different words initially starts from $N$
and slowly begins to decay [Fig.~S\ref{fig_S1}(a).] The relative
frequencies of the final configurations with one, two, and three
words are very similar [Fig.~2(b) of the main article vs
Fig.~S\ref{fig_S1}(b)], and the underlying opinion clusters
(communities) exhibited this way are the same. Note that for the
one-word-per-agent initial conditions, out of $10,000$ independent
runs, we also recorded 3 runs with four surviving opinion clusters
($N_{d}$$=$$4$) after $10^4$ time steps, corresponding to a $3\times
10^{-4}$ relative frequency. Since it is three orders of magnitudes
smaller than the probability of other possible final configurations,
it is not visible on the same scales and is not shown in
Fig.~S\ref{fig_S1}(b).

\section*{The Naming Game with finite-memory agents}

%
As was discussed by Dall'Asta et al (2006), the typical memory need
(the maximum number of words in its list at any given time) of an
agent of degree $k$ is of $\sqrt{k}$. Limiting the agents' memory to
a finite value $L$, in general, can slow down the consensus process
(Wang et al, 2007). For the particular friendship network we used
here, $k_{\rm min}=1$, $k_{\rm max}=33$, and $\overline{k}=8.8$.
Here we considered a ``first in - first out" finite-memory version
of the NG: in case of an unsuccessful communication, if the memory
of the listening agents is full, it drops the word from its list
which has been there the longest, and adds the one just heard. All
the other rules of the NG, as described in Sec. 2 of the main
article, remain the same.

\begin{figure}[t]
\vspace*{3.0truecm}
  \includegraphics{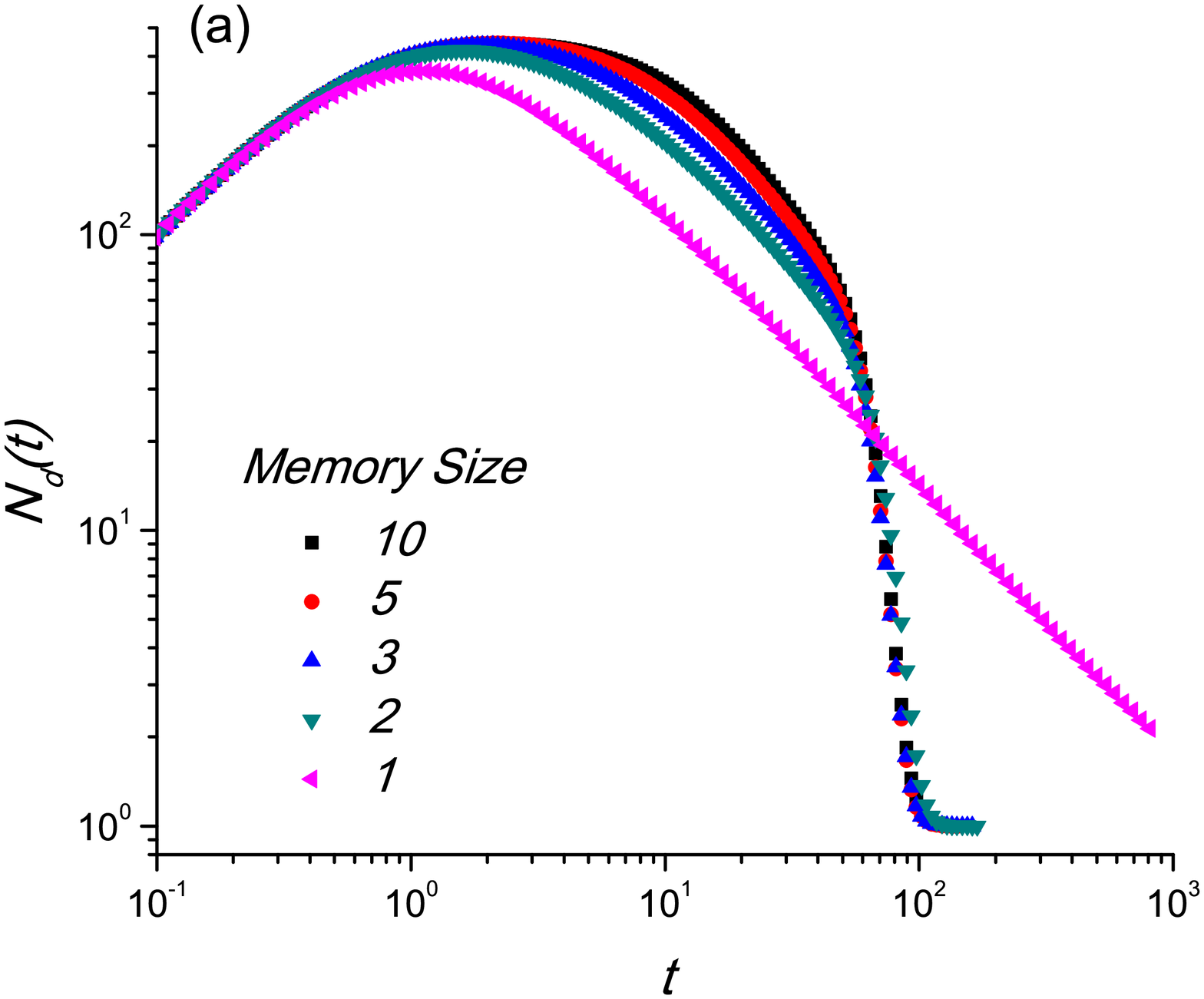}
  \includegraphics{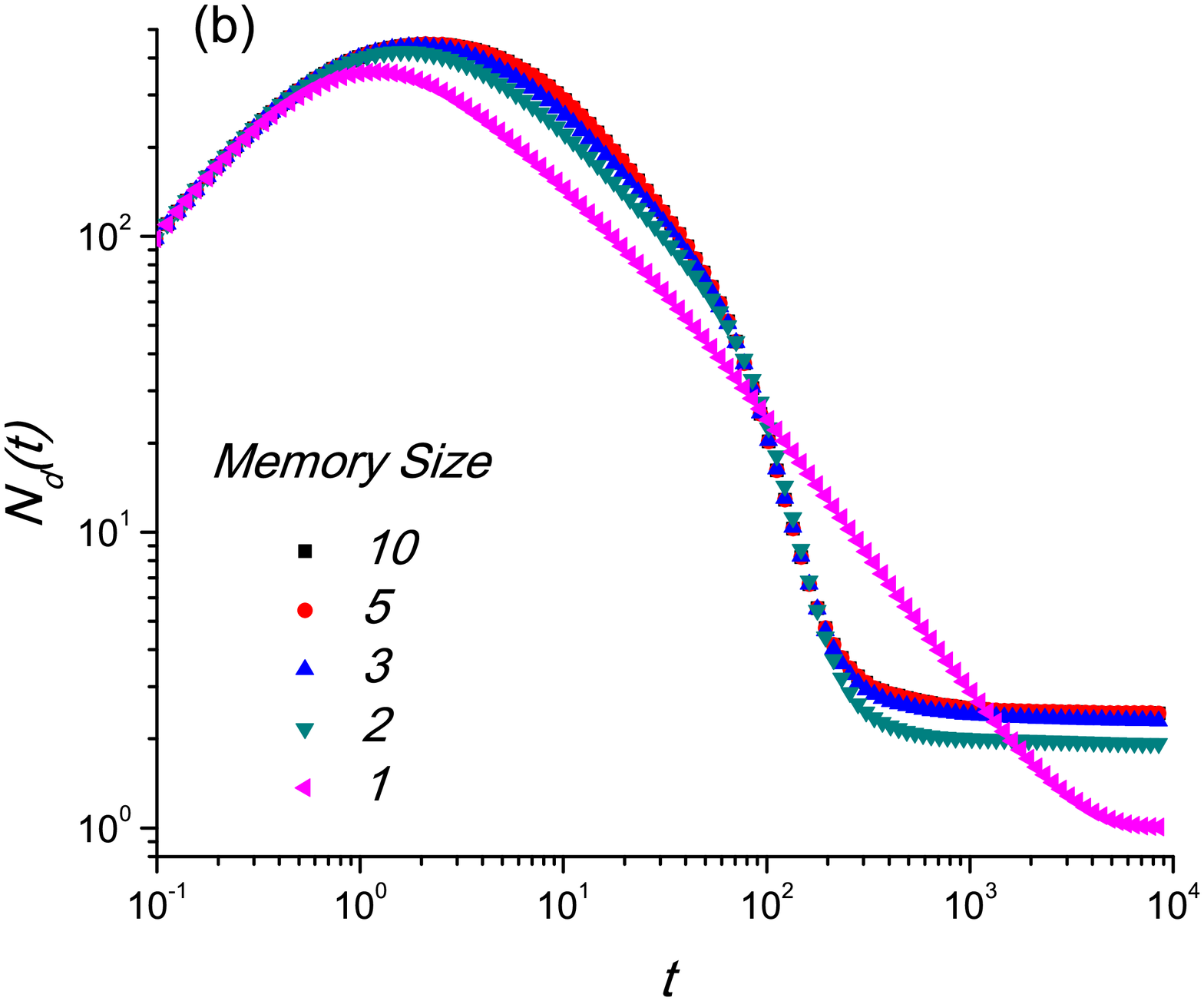}
\vspace*{3.5truecm}
\caption{Naming Game with finite-memory agents. The number of
different words vs time on (a) the Watts-Strogatz  network; (b) a
friendship network with the same number of nodes, average degree,
and clustering coefficients ($N=1,127$, $\overline{k}=8.8$, and
$\overline{C}=0.067$). All curves represent averages over $10,000$
independent runs.}
\label{fig_S2}
\end{figure}

For comparison, here we show the agreement process in terms of
number of different words (communities) vs time for both the
Watts-Strogatz network and for the friendship network  with the same
number of nodes, average degree, and clustering coefficient
[Fig.~S\ref{fig_S2}]. Since the average degree of these networks is
rather small, there is very little variation in the late-stage
agreement process and convergence times (the infinite-memory
behavior is asymptotically approached as the memory length is of
order $\sqrt{\bar{k}}$). The only exception is the $L$$=$$1$ case:
here the listener simply replaces its current opinion with the one
of the speaker's, hence the model becomes equivalent to the
$q$-state voter model (with $q=\infty$) (Howard and Godr\`eche,
1998). Unlike the NG with $L$$\geq$$2$, the voter model on random
graphs has no tendency to form compact domains; the ``interfaces" or
boundaries separating domains stochastically disintegrate. While the
voter model on finite networks eventually orders, it is the result
of a large spontaneous fluctuation of ordered regions spanning the
full system (Castellano et al, 2003; Castellano et al, 2005). Global
consensus for the NG with $L$$=$$1$ (equivalent to the voter model)
on networks with no community structure takes much longer than for
$L$$\geq$$2$ [Fig.~S\ref{fig_S2}(a)]. On networks with community
structure (such as the friendship network), however, the tendency of
the NG with $L$$\geq$$2$ to form domains leads to the formation of
compact long-living or meta-stable opinion clusters, hence global
order is rarely reached. On the other hand, when $L$$=$$1$, there
are no stable domain boundaries and the underlying community
structure has little or no effect on reaching global consensus
[Fig.~S\ref{fig_S2}(b)]. The NG with L=1 (voter model) orders on any
finite network, essentially independently of the underlying
community structure.

\section*{References}

%

\vspace*{0.1truecm} \noindent Castellano C, Vilone D, Vespignani A
(2003) Incomplete ordering of the voter model on small-world
networks. Europhys. Lett. 63:153--158

\vspace*{0.1truecm} \noindent Castellano C, Loreto V, Barrat A,
Cecconi F, Parisi D (2005) Comparison of voter and Glauber ordering
dynamics on networks. Phys. Rev. E 71:066107

\vspace*{0.1truecm} \noindent Dall'Asta L, Baronchelli A, Barrat A,
Loreto V (2006) Nonequilibrium dynamics of language games on complex
networks. Phys. Rev. E 74:036105

\vspace*{0.1truecm} \noindent Howard M, Godr\`eche C (1998)
Persistence in the Voter model: continuum reaction-diffusion
approach. J. Phys. A: Math. Gen. 31:L209--L215

\vspace*{0.1truecm} \noindent Wang WX, Lin BY, Tang CL, Chen GR
(2007) Agreement dynamics of finite-memory language games on
networks. Eur. Phys. J. B 60:529--536


\end{document}